\documentstyle[amstex,prd,aps,epsfig]{revtex}

\newcommand{\be}{\begin{equation}}
\newcommand{\ee}{\end{equation}}
\newcommand{\ba}{\begin{eqnarray}}
\newcommand{\ea}{\end{eqnarray}}

\begin{document}
\draft
\title{Electron-Electron Bound States in Maxwell-Chern-Simons-Proca QED$_{3}$ }
\author{H. Belich$^{a,b}$, O.M. Del Cima$^{a}$, M.M. Ferreira Jr.$^{a,c}$ and J.A.
Helay\"{e}l-Neto$^{a,b}$\thanks{{\tt e-mails: belich\copyright cbpf.br,
delcima\copyright gft.ucp.br, manojr\copyright cbpf.br, helayel\copyright
gft.ucp.br}}}
\address{$^{a}${\it Grupo de F\'{i}sica Te\'{o}rica Jos\'{e} Leite Lopes}\\
Petr\'{o}polis - RJ - Brazil\\
$^{b}${\it Centro Brasileiro de Pesquisas F\'{i}sicas (CBPF)},\\
Coordena\c{c}\~{a}o de Teoria de Campos e Part\'{i}culas (CCP), \\
Rua Dr. Xavier Sigaud, 150 - Rio de Janeiro - RJ 22290-180 - Brazil.\\
$^{c}${\it Universidade Federal do Maranh\~{a}o (UFMA)}, \\
Departamento de F\'{i}sica, Campus Universit\'{a}rio do Bacanga, \\
S\~{a}o Luiz - MA, 65085-580 - Brazil. }
\maketitle

\begin{abstract}
We start from a parity-breaking MCS QED$_{3}$ model with spontaneous
breaking of the gauge symmetry as a framework for evaluation of the
electron-electron interaction potential and for attainment of numerical
values for the $e^{-}e^{-}-$ bound state. Three expressions ($V_{\text{eff}%
_{\uparrow \uparrow }},V_{\text{eff}_{\uparrow \downarrow }},V_{\text{eff}%
_{\downarrow \downarrow }})$ are obtained according to the polarization
state of the scattered electrons. In an energy scale compatible with
Condensed Matter electronic excitations, these three potentials become
degenerated. The resulting potential is implemented in the Schr\"{o}dinger
equation and the variational method is applied to carry out the electronic
binding energy. The resulting binding energies in the scale of $10-100$ $meV$
and a correlation length in the scale of $10-30$\AA\ are possible
indications that the MCS-QED$_{3}$ model adopted may be suitable to address
an eventual case of $e^{-}e^{-}$ pairing in the presence of parity-symmetry
breakdown. The data analyzed here suggest an energy scale of $10$-$100$ $meV$
to fix the breaking of the $U(1)$-symmetry.
\end{abstract}

\pacs{PACS numbers: 11.10.Kk 11.15.Ex 74.20.Mn\hspace{4cm}ICEN-PS-01/17}

\section{Introduction}

The advent of high-T$_{c}$ superconductivity \cite{Bednorz}, in 1986,
brought about a great excitation in both the theoretical and experimental
physical panorama, drawing attention for the issue of formation of Cooper
pairs in planar systems. In the late 90%
\'{}%
s, there arose a field-theoretical approach to address the mechanism of
electronic pairing: the evaluation of the electron-electron M\"{o}ller
scattering as a tool for the attainment of the $e^{-}e^{-}$ interaction
potential in the nonrelativistic approximation. This line of action searches
for an attractive potential in such a way to induce the formation of
correlated electron-electron pairs, (the charge carriers of the high-T$_{c}$
superconductors). The present work shall follow this general procedure.

By direct application of the Gauss%
\'{}%
s law in ($1+2)$-dimensions for the massless gauge field, the Coulombian
interaction takes on the form of a confining potential $\left( \ln r\right) $%
. The Kato condition \cite{Chadan} establishes the finiteness of the number
of bound states, in $D=1+2,$ associated to a certain potential $V,$ and can
be used as a criterion for determining the character confining or
condensating of the potential. The fact the logarithmic potential to be
confining (according to the Kato criterion) indicates it does not lead to
bound states, becoming clear the need of a finite range, screened
interaction. The Chern-Simons (CS)\ term \cite{DJ} is then introduced as the
generator of (topological) mass for the photon, implying an intensive
screening of the Coulombian interaction. The Maxwell-Chern-Simons (MCS)
model, a particular case of Planar Quantum Electrodynamics - QED$_{3},$ then
arose as a theoretical framework able for providing an attractive but not
confining electron-electron interaction. This model was then used by some
authors \cite{Kogan}, \cite{Girotti}, \cite{Dobroliubov}, \cite{Groshev} as
basic tool for evaluation of the M\"{o}ller scattering amplitude at
tree-level, whose Fourier transform (in the Born approximation) yields the $%
e^{-}e^{-}$ interaction potential. In a general way, these works have led to
the same result:\ the electron-electron potential comes out attractive
whenever the topological mass $\left( \vartheta \right) $ exceeds the
electron mass $\left( m_{e}\right) $. Georgelin and Wallet \cite{Georgelin}
started from two MCS-QED$_{3}$ Lagrangians, the first (second) with the
gauge field nonminimally coupled to fermions (bosons), in such a way to
consider the introduction of the anomalous momentum of the electron in the
problem. Working in the perturbative regime ($1/k\ll 1),$ these authors
found an attractive potential for fermions $\left( V_{\psi \psi }<0\right) ,$
and also for scalar bosons $\left( V_{\varphi \varphi }<0\right) ,$ in the
nonrelativistic approximation. The presence of the nonminimal coupling seems
to be the key-factor for the attainment of the attractive potential between
charges with the same sign. In this case, the potential remains negative
even in the limit of a small topological mass $\left( \vartheta \ll
m_{e}\right) $, under a suitable choice of parameters. The
nonrenormalizability of this model (due to the nonminimal coupling),
however, implies a restriction to the validity of their results only at
tree-level calculations.

All the MCS models, except the one exposed in Ref. \cite{Georgelin}, failed
under the perspective of yielding a realistic electron-electron condensation
into the domain of a Condensed Matter system due to the condition $\vartheta
>m_{e},$ necessary for making the $e^{-}e^{-}$ pairing feasible. One must
believe to be unlikely the existence of a physical excitation with so large
energy in a real solid state system (the superconductors usually are
characterized by excitations\ in the $meV$ scale). We will see that the
introduction of the Higgs mechanism in the context of the
MCS-Electrodynamics will bring out a negative contribution to the scattering
potential that will allow a global attractive potential despite the
condition $\vartheta >m_{e}$.

In our work, we shall rely on a version of planar QED for which the photonic
excitations appear as a by-product of a spontaneous symmetry breaking (SSB)
realized on the MCS Lagrangian. The consideration of a Higgs sector (a
complex scalar field endowed with a self-interaction potential so as to
induce a SSB) in the context of the MCS model provides a new mass term to
the topological gauge field: the well-known Proca term $\left( m^{2}A_{\mu
}A^{\mu }\right) $. In this way, once the spontaneous breaking of the local
U(1)-symmetry has taken place, a neutral massive Higgs scalar remains and
the gauge field becomes a Maxwell-Chern-Simons-Proca vector field, a clear
reference to its two mass components (the topological and the Proca one).
The physical mass of such a photon, that may assume two different values,
will be written in terms of these two mass parameters, as explicitly given
by the expressions read off from the poles of the gauge-field propagator
(see Section III). For our purposes, one can assert that the enhancement of
complexity determined by the coexistence of a topological and a Proca term
in the gauge sector is compensated by the attainment of a gauge propagator
with two massive poles (standing for the photon mass). Operationally, in the
perspective of a tree-level field-theory investigation, the determination of
the gauge propagator and the Feynman rules enable us to derive the
interaction potential between two elementary particles as mediated by this
gauge field. This paper, therefore, adopts as starting point a MCS-Proca
Lagrangian with the clear purpose of performing a usual field-theory
derivation (in the non-relativistic limit) of an interaction potential,
which is further on applied to obtain bound states (in a typical quantum
mechanical procedure). Last but not least, the fact that the photon becomes
massive is a microscopical information that renders feasible the observation
of the Meissner effect in such a system, which opens the applicability of
such kind of model for an \ eventual superconducting planar system endowed
with parity breaking \cite{Parity-breaking}. This theoretical possibility,
however, is out of the scope of this work.

In a recent paper \cite{Int-Journal}, we have derived an interaction
potential associated to the scattering of two identically polarized
electrons in the framework of a Maxwell-Chern-Simons QED$_{3}$ with
spontaneous breaking of local-U(1) symmetry. Our result revealed the
interesting possibility of an attractive electron-electron interaction
whenever the contribution stemming from the Higgs sector overcomes the
repulsive contribution from the gauge sector, which can be achieved by an
appropriate fitting of the free parameters. In the present work, we
generalize the results attained in Ref. \cite{Int-Journal}, \cite{Tese}
contemplating the existence of two fermionic families $\left( \psi _{+},\psi
_{-}\right) ,$ and performing the numerical evaluation of the $e^{-}e^{-}$
binding energies. The procedure here accomplished is analogous to the one
enclosed in Ref. \cite{Int-Journal}, \cite{Tese}: starting from a QED$_{3}$
Lagrangian (now built up by two spinor polarizations, $\psi _{+},\psi _{-})$
with SSB, one evaluates the M\"{o}ller scattering amplitudes (in the
nonrelativistic approximation) having the Higgs and the massive photon as
mediators and the corresponding interaction potential, that now emerges in
three different expressions:\ $V_{_{\uparrow \uparrow }},V_{_{\uparrow
\downarrow }},V_{_{\downarrow \downarrow }}$ (depending on the spin
polarization of the scattered electrons). The same theoretical possibility
of attractiveness, pointed out in Ref. \cite{Int-Journal}, is now manifested
by these three potentials. A numerical procedure (variational method) is
then implemented in order to carry out the binding energy of the Cooper
pairs. Having in mind the nonrelativistic approximation, a reduced potential
is implemented into the Schr\"{o}dinger equation, whose numerical solution
provides the data contained in Tables \ref{table1}, \ref{table2}, \ref
{table3}. The achievement of binding energies in the $meV$ scale and
correlation length in the $10-30$\AA\ scale is an indicative that the
adopted MCS-QED$_{3}$ model may be suitable for addressing an eventual
electronic pairing in a system endowed with parity-breaking.

This paper is outlined as follows: in Section II, we present the QED$_{3}$
Lagrangian, its general features and one realizes the spontaneous breaking
of U(1)-local symmetry that generates the Higgs boson and the
Maxwell-Chern-Simons-Proca photon; in Section III, one evaluates the
amplitudes for the M\"{o}ller scattering; their Fourier transform will
provide the $e^{-}e^{-}$ interaction potentials $V_{_{\uparrow \uparrow
}},V_{_{\uparrow \downarrow }},V_{_{\downarrow \downarrow }}$ (despite the
complex form of these potentials, they maintain the theoretical possibility
of being attractive); in Section IV,\ one performs an analysis in order to
obtain the $e^{-}e^{-}$ binding energies by means of the numerical solution
of the Schr\"{o}dinger equation (by the variational method), whose results
are disposed in Tables \ref{table1}, \ref{table2}, \ref{table3}. In Section
V, we present our General Conclusions.

.

\section{The MCS QED$_{3}$ with Spontaneous Symmetry Breaking and two Spinor
Polarizations}

The action for a QED$_{3}$ model built up by two polarization fermionic
fields ($\psi _{+},\psi _{-}$), a gauge $\left( A_{\mu }\right) $ and a
complex scalar field $\left( \varphi \right) $, mutually coupled, and
endowed with spontaneous breaking of a local U(1)-symmetry \cite{N.Cimento}, 
\cite{Int-Journal}, reads as

\begin{align}
S_{QED-MCS} & = \int d^{3}x\{-\frac{1}{4}F^{\mu \nu }F_{\mu \nu }+i\overline{%
\psi }_{+}\gamma ^{\mu }D_{\mu }\psi _{+}+i\overline{\psi }_{-}\gamma ^{\mu
}D_{\mu }\psi _{-}+%
{\frac12}%
\theta \epsilon ^{\mu v\alpha }A_{\mu }\partial _{v}A_{\alpha }-m_{e}(%
\overline{\psi }_{+}\psi _{+}-\overline{\psi }_{-}\psi _{-})+  \nonumber \\
& - y(\overline{\psi }_{+}\psi _{+}-\overline{\psi }_{-}\psi _{-})\varphi
^{\ast }\varphi +D^{\mu }\varphi ^{\ast }D_{\mu }\varphi -V(\varphi ^{\ast
}\varphi )\},  \label{actionMCS}
\end{align}
where $V(\varphi ^{\ast }\varphi )$ represents the sixth-power
self-interaction potential,

\begin{equation}
V(\varphi ^{\ast }\varphi )=\mu ^{2}\varphi ^{\ast }\varphi +\frac{\zeta }{2}%
(\varphi ^{\ast }\varphi )^{2}+\frac{\lambda }{3}(\varphi ^{\ast }\varphi
)^{3},
\end{equation}
which is responsible for the SSB; it is the most general one renormalizable
in $1+2$ dimensions \cite{Delcima}. The mass dimensions of the parameters $%
\mu ,\zeta ,\lambda $ and $y$ are respectively: 1,1,0 and 0. For the present
purpose, we are interested only on stable vacuum, restriction satisfied by
imposing some conditions on the potential parameters: $\lambda >0,\zeta <0$
and $\mu ^{2}\leq \frac{3\zeta ^{2}}{16\lambda }.$ The covariant derivatives
are defined as: $D_{\mu }\psi _{\pm }=(\partial _{\mu }+ie_{3}A_{\mu })\psi
_{\pm }$ \ and $D_{\mu }\varphi =(\partial _{\mu }+ie_{3}A_{\mu })\varphi ,$
where $e_{3}$ is the coupling constant of the $U(1)$-local gauge symmetry,
here with dimension of (mass)$^{1/2}$, particularity that will be more
explored in the numerical analysis section. In $\left( 1+2\right) -$%
dimensions, a fermionic field has its spin polarization fixed up by the mass
sign \cite{Binegar}; however, in the action (\ref{actionMCS}), it is
manifest the presence of two spinor fields of opposite polarization. In this
sense, it is necessary to stress that we have two positive-energy spinors
(two spinor families), both solutions of the Dirac equation, each one with
one polarization state according to the sign of the mass parameter, instead
of the same spinor with two possibilities of spin-polarization.

Considering $\langle \varphi \rangle =v,$ the vacuum expectation value for
the scalar field product $\varphi ^{\ast }\varphi $ is given by: 
\[
\langle \varphi ^{\ast }\varphi \rangle =v^{2}=-\zeta /\left( 2\lambda
\right) +\left[ \left( \zeta /\left( 2\lambda \right) \right) ^{2}-\mu
^{2}/\lambda \right] ^{1/2}, 
\]
while the condition for minimum reads as: $\mu ^{2}+\frac{\zeta }{2}%
v^{2}+\lambda v^{4}=0$. \ After the spontaneous symmetry breaking, the
scalar complex field can be parametrized by $\varphi =v+H+i\theta $, where $%
H $ represents the Higgs scalar field and $\theta $ the would-be Goldstone
boson; the SSB\ will be manifest when this parametrization is replaced in
the action (\ref{actionMCS}). Thereafter, in order to preserve the manifest
renormalizability of the model, one adopts the 
\'{}%
t Hooft gauge by adding \ the fixing gauge term $\left( S_{R_{\xi
}}^{gt}=\int d^{3}x[-\frac{1}{2\xi }(\partial ^{\mu }A_{\mu }-\sqrt{2}\xi
M_{A}\theta )^{2}]\right) $ to the broken action; finally, by retaining only
the bilinear and the Yukawa interaction terms, one has,

\begin{align}
{S}_{{\rm QED}}^{{\rm SSB}}& =\int d^{3}x\biggl\{-\frac{1}{4}F^{\mu \nu
}F_{\mu \nu }+\frac{1}{2}M_{A}^{2}A^{\mu }A_{\mu }-\frac{1}{2\xi }(\partial
^{\mu }A_{\mu })^{2}+\overline{\psi }_{+}(i{\rlap{\hbox{$\mskip 1 mu /$}}%
\partial }-m_{eff})\psi _{+}+\overline{\psi }_{-}(i{%
\rlap{\hbox{$\mskip 1 mu
/$}}\partial }+m_{eff})\psi _{-}+%
{\frac12}%
\theta \epsilon ^{\mu v\alpha }A_{\mu }\partial _{v}A_{\alpha }+  \nonumber
\\
& +\partial ^{\mu }H\partial _{\mu }H-M_{H}^{2}H^{2}+\partial ^{\mu }\theta
\partial _{\mu }\theta -M_{\theta }^{2}\theta ^{2}-2yv(\overline{\psi }%
_{+}\psi _{+}-\overline{\psi }_{-}\psi _{-})H-e_{3}\left( \overline{\psi }%
_{+}{\rlap{\hbox{$\mskip 1 mu /$}}A}\psi _{+}+\overline{\psi }_{-}{%
\rlap{\hbox{$\mskip 1 mu /$}}A}\psi _{-}\right) \biggr\},  \label{actionMCS3}
\end{align}
whose mass parameters, 
\begin{equation}
M_{A}^{2}=2v^{2}e_{3}^{2},\text{ \ \ \ \ }m_{eff}=m_{e}+yv^{2},\ \ \ M_{%
{\small H}}^{2}=2v^{2}(\zeta +2\lambda v^{2}),\text{ \ }M_{\theta }^{2}=\xi
M_{A}^{2},
\end{equation}
are entirely or partially dependent on the SSB mechanism. The Proca mass, $%
M_{A}^{2}$, represents the mass acquired by the photon through the Higgs
mechanism, while the Higgs mass, $M_{H}^{2}$, is the one associated with the
real scalar field. The Higgs mechanism also corrects the mass of the
electron, resulting in an effective electronic mass, \ $m_{eff}$. On the
other hand, the would-be Goldstone mode, endowed with mass $(M_{\theta
}^{2}) $, does not represent a physical excitation, since $\xi $\ is just a
unphysical (dimensionless) gauge-fixing parameter. At this moment, it is
instructive to point out the presence of two photon mass-terms in eq. (\ref
{actionMCS3}): the Proca and the topological one. The physical mass of the
gauge field will emerge as a function of two mass parameters, as shown in
the next Section.

\section{The Electron-Electron Scattering Potential in the nonrelativistic
Limit}

In the low-energy limit (Born approximation), the two-particle interaction
potential is given by the Fourier transform of the two-particle scattering
amplitude \cite{Sakurai}. \ It is important to stress that, in the case of
the nonrelativistic M\"{o}ller scattering, one should consider only the
t-channel (direct scattering) \cite{Sakurai} even for indistinguishable
electrons, since in this limit they recover the classical notion of
trajectory. The M\"{o}ller scattering will be\ mediated by two particles:
the Higgs scalar and the massive gauge field. From the action (\ref
{actionMCS3}), one reads off the propagators associated to the Higgs scalar
and Maxwell-Chern-Simons-Proca field:

\begin{eqnarray}
\text{\ }\langle H(k)H(-k)\rangle &=&\frac{i}{2}\frac{1}{k^{2}-M_{H}^{2}};%
\text{ \ \ \ }\langle A_{\mu }(k)A_{\nu }(-k)\rangle =-i\biggl\{\frac{%
k^{2}-M_{A}^{2}}{(k^{2}-M_{A}^{2})^{2}-k^{2}\theta ^{2}}\biggl(\eta _{\mu
\nu }-\frac{k_{\mu }k_{\nu }}{k^{2}}\biggr)+  \nonumber \\
&&+\frac{\xi }{(k^{2}-\xi M_{A}^{2})}\frac{k_{\mu }k_{\nu }}{k^{2}}+\frac{%
\theta }{(k^{2}-M_{A}^{2})^{2}-k^{2}\theta ^{2}}i\epsilon _{\mu \alpha \nu
}k^{\alpha }\biggr\}.
\end{eqnarray}
The photon propagator can be split in the following form,

\[
\langle A_{\mu }A_{\nu }\rangle =-i\left[ \frac{C_{+}}{k^{2}-M_{+}^{2}}+%
\frac{C_{-}}{k^{2}-M_{-}^{2}}\right] (\eta _{\mu \nu }-\frac{k_{\mu }k_{\nu }%
}{k^{2}})+\frac{-i\xi k_{\mu }k_{\nu }}{k^{2}(k^{2}-\xi M_{A}^{2})}+i\left[ 
\frac{C}{k^{2}-M_{+}^{2}}-\frac{C}{k^{2}-M_{-}^{2}}\right] \epsilon _{\mu
\alpha \nu }k^{\alpha }, 
\]
with the positive definite constants $C_{+},C_{-},C$ and the quadratic
masses poles $M_{+}^{2}$ and $M_{-}^{2}$ given by:

\begin{equation}
C_{\pm }=\frac{1}{2}\left[ 1\pm \frac{\theta }{\sqrt{4M_{A}^{2}+\theta ^{2}}}%
\right] ;\text{ \ \ }C=\frac{1}{\sqrt{4M_{A}^{2}+\theta ^{2}}};\text{ \ }%
M_{\pm }^{2}=\frac{1}{2}\left[ (2M_{A}^{2}+\theta ^{2})\pm |\theta |\sqrt{%
4M_{A}^{2}+\theta ^{2}}\right] .
\end{equation}
Here, $C_{\pm }$\ and $C$\ are constants with mass dimension $0$\ and $-1$
respectively, whereas $M_{\pm }^{2}$\ represent the two possible expressions
for the physical mass of the photon (around which occur photonic
excitations). Consequently, these two masses, rather than \ $M_{A}^{2}$\ and 
$\theta ^{2}$, will be the relevant ones in the forthcoming evaluation of
the interaction potential.

From the action (\ref{actionMCS3}), it is easy to extract the vertex Feynman
rules: $V_{\psi _{\pm }H\psi _{\pm }}=\pm 2ivy;V_{\psi A\psi }=ie_{3}\gamma
^{\mu }.$ Since in the low-energy limit only the t-channel must be
considered, the whole scattering amplitudes are written in the form:

\begin{eqnarray}
-i{\cal M}_{\pm H\pm } &=&\overline{u}_{\pm }(p_{1})(\pm 2ivy)u_{\pm
}(p_{1}^{^{\prime }})\left[ \langle H(k)H(-k)\rangle \right] \overline{u}%
_{\pm }(p_{2})(\pm 2ivy)u_{\pm }(p_{2}^{^{\prime }}), \\
-i{\cal M}_{\pm H\mp } &=&\overline{u}_{\pm }(p_{1})(\pm 2ivy)u_{\pm
}(p_{1}^{^{\prime }})\left[ \langle H(k)H(-k)\rangle \right] \overline{u}%
_{\mp }(p_{2})(\mp 2ivy)u_{\mp }(p_{2}^{^{\prime }}), \\
-i{\cal M}_{\pm A\pm } &=&\overline{u}_{\pm }(p_{1})(ie_{3}\gamma ^{\mu
})u_{\pm }(p_{1}^{^{\prime }})\left[ \langle A_{\mu }(k)A_{\nu }(-k)\rangle %
\right] \overline{u}_{\pm }(p_{2})(ie_{3}\gamma ^{\nu })u_{\pm
}(p_{2}^{^{\prime }}), \\
-i{\cal M}_{\pm A\mp } &=&\overline{u}_{\pm }(p_{1})(ie_{3}\gamma ^{\mu
})u_{\pm }(p_{1}^{^{\prime }})\left[ \langle A_{\mu }(k)A_{\nu }(-k)\rangle %
\right] \overline{u}_{\mp }(p_{2})(ie_{3}\gamma ^{\nu })u_{\mp
}(p_{2}^{^{\prime }}).
\end{eqnarray}
The first two expressions represent the scattering amplitude mediated by the
Higgs particles for equal and opposite electron polarizations, while in the
last two ones the mediator is the massive Chern-Simons-Proca photon. The
spinors $u_{+}(p),$ $u_{-}(p)$ stand for the positive-energy solution of the
Dirac equation, satisfying the normalization conditions: $\overline{u}_{\pm
}(p)u_{\pm }(p)=\pm 1.$ Working in the center-of-mass frame, the momenta of
the interacting particles and the momentum transfer take a simpler form,
useful for writing the spinors $u_{+}(p),$ $u_{-}(p)$, as it is properly
shown in the Appendix. With these definitions, one carries out the fermionic
current elements, \ also displayed in the Appendix, so that the evaluation
of the scattering amplitudes (for low momenta approximation), at tree-level,
associated to the Higgs and the gauge particle become: 
\begin{equation}
\text{ \ \ \ \ \ \ }{\cal M}_{Higgs}=-2v^{2}y^{2}\biggl(\frac{1}{%
\overrightarrow{k}^{2}+M_{{\small H}}^{2}}\biggr),  \label{MHiggs}
\end{equation}

\[
{\cal M}_{\uparrow A\uparrow }={\cal M}_{1}+{\cal M}_{2}+{\cal M}_{3},\text{%
\ \ }{\cal M}_{\downarrow A\downarrow }={\cal M}_{1}-{\cal M}_{2}+{\cal M}%
_{3},\text{ \ \ \ }{\cal M}_{\uparrow A\downarrow }={\cal M}_{\downarrow
A\uparrow }={\cal M}_{1}+{\cal M}_{3}, 
\]
with:

\begin{equation}
{\cal M}_{1}=e_{3}^{2}\left[ \frac{C_{+}}{\overrightarrow{k}^{2}+M_{+}^{2}}+%
\frac{C_{-}}{\overrightarrow{k}^{2}+M_{-}^{2}}\right] ,\text{ }{\cal M}_{2}=%
\frac{e_{3}^{2}\overrightarrow{k}^{2}}{m_{\text{{\small eff}}}}\left[ \frac{C%
}{\overrightarrow{k}^{2}+M_{+}^{2}}-\frac{C}{\overrightarrow{k}^{2}+M_{-}^{2}%
}\right] ,\text{ }{\cal M}_{3}=\frac{-i\sin \phi }{(1-\cos \phi )}{\cal M}%
_{2},
\end{equation}
where it was used $\overrightarrow{k}^{2}=2p^{2}(1-\cos \phi )$.
Furthermore, it is clear that the Higgs amplitude is independent of the
electron polarization, while the gauge amplitude splits into three different
expressions, depending on the polarization of the scattered electrons. The
terms ${\cal M}_{1}$,${\cal M}_{2}$ correspond to the real part of the
M\"{o}ller scattering amplitude, while ${\cal M}_{3}$ describes the
Aharonov-Bohm amplitude for fermions \cite{Kogan},\cite{Dobroliubov},\cite
{Georgelin}. The interaction potentials are obtained through the Fourier
transform of the scattering amplitude (inside the Born approximation limit): 
$V(\overrightarrow{r})=\int \frac{d^{2}k}{(2\pi )^{2}}{\cal M}e^{i%
\overrightarrow{k}.\overrightarrow{r}}.$ According to this approximation,
Eq.(\ref{MHiggs}) yields an attractive Higgs potential,

\begin{equation}
V_{Higgs}(r)=-\frac{1}{2\pi }2v^{2}y^{2}K_{0}(M_{{\small H}}r),\text{ \ \ }
\end{equation}
while in the gauge sector there appear three different potentials (depending
on the polarization state):

\[
\text{\ }V_{gauge\text{ }\uparrow \uparrow }(r)=V_{1}(r)+V_{2}(r)+V_{3}(r),%
\text{ \ }V_{gauge\text{ }\uparrow \downarrow }(r)=V_{1}(r)+V_{3}(r),\text{
\ }V_{gauge\text{ }\downarrow \downarrow }(r)=V_{1}(r)-V_{2}(r)+V_{3}(r), 
\]
$V_{1}(r),$ $V_{2}(r),$ $V_{3}(r)$ being respectively the Fourier transforms
of the amplitudes ${\cal M}_{1,}{\cal M}_{2},{\cal M}_{3}$, given explicitly
by:

\begin{align}
V_{1}(r)& =\frac{e_{3}^{2}}{2\pi }\biggl[%
C_{+}K_{0}(M_{+}r)+C_{-}K_{0}(M_{-}r)\biggr],\text{ } \\
V_{2}(r)& =-\frac{e_{3}^{2}}{2\pi }\frac{C}{m_{\text{{\tiny eff}}}}\biggl[%
M_{+}^{2}K_{0}(M_{+}r)-M_{-}^{2}K_{0}(M_{-}r)\biggr], \\
V_{3}(r)& =2\frac{e_{3}^{2}}{2\pi }\frac{Cl}{m_{\text{{\tiny eff}}}r}\biggl [%
M_{+}K_{1}(M_{+}r)-M_{-}K_{1}(M_{-}r)\biggr].
\end{align}

The complete potential expressions are obtained joining the Higgs and gauge
contributions: $V(r)=V_{Higgs}+V_{gauge}$:

\begin{align}
V(r)_{\uparrow \uparrow }& =-\frac{1}{2\pi }2v^{2}y^{2}K_{0}(M_{h}r)+\frac{%
e_{3}^{2}}{2\pi }\text{ }\biggl\{(C_{+}-\frac{C}{m}%
M_{+}^{2})K_{0}(M_{+}r)+(C_{-}+\frac{C}{m_{\text{{\tiny eff}}}}%
M_{-}^{2})K_{0}(M_{-}r)+  \nonumber \\
& +2\frac{Cl}{m_{_{\text{{\tiny eff}}}}r}%
(M_{+}K_{1}(M_{+}r)-M_{-}K_{1}(M_{-}r))\biggr\},  \label{V1}
\end{align}

\begin{align}
V(r)_{\uparrow \downarrow }& =-\frac{1}{2\pi }2v^{2}y^{2}K_{0}(M_{h}r)+\frac{%
e_{3}^{2}}{2\pi }\text{ }\biggl\{C_{+}K_{0}(M_{+}r)+C_{-}K_{0}(M_{-}r)+2%
\frac{Cl}{m_{\text{{\tiny eff}}}r}[M_{+}K_{1}(M_{+}r)+  \nonumber \\
& -M_{-}K_{1}(M_{-}r)]\biggr\},  \label{V2}
\end{align}

\begin{align}
V(r)_{\downarrow \downarrow }& =-\frac{1}{2\pi }2v^{2}y^{2}K_{0}(M_{h}r)+%
\frac{e_{3}^{2}}{2\pi }\text{ }\biggl\{(C_{+}+\frac{C}{m_{\text{{\tiny eff}}}%
}M_{+}^{2})K_{o}(M_{+}r)+(C_{-}-\frac{C}{m_{\text{{\tiny eff}}}}%
M_{-}^{2})K_{0}(M_{-}r)  \nonumber \\
& +2\frac{Cl}{m_{\text{{\tiny eff}}}r}(M_{+}K_{1}(M_{+}r)-M_{-}K_{1}(M_{-}r))%
\biggr\}.  \label{V3}
\end{align}

Here, $K_{0}(x)$ and $K_{1}(x)$ are the modified Bessel functions and $l$ is
the angular momentum. The last three equations represent the tree-level
potentials evaluated at the Born approximation. Now, it is convenient to
define the limit of validity of the potentials (\ref{V1}), (\ref{V2}), (\ref
{V3}). They have been derived in the low-energy limit, consequently they
must be valid in the perturbative regime, where the loop corrections are
negligible before the semi-classical terms. For a typical MCS model, the
perturbative limit is given by $\frac{e^{2}}{\theta }\ll 1$; in the case of
the present model, nevertheless, there are four dimensionless parameters \ ${%
e_{3}^{2}}/m$, ${e_{3}^{2}}/{M_{H}}$, ${e_{3}^{2}}/{M_{+}}$, ${e_{3}^{2}}/{%
M_{-}}$. According to the discussion realized in Ref. \cite{Int-Journal},
the pertubative regime is valid whenever ${e_{3}^{2}}/{M_{+}}\ll 1$ and $%
y\ll 1$ (the first condition obviously implies ${e}_{3}^{{2}}/m\ll 1$).

A remarkable point to be highlighted concerns the attainment of three
different potentials: $V(r)_{\uparrow \uparrow },V(r)_{\uparrow \downarrow
},V(r)_{\downarrow \downarrow }$. Our results put in explicit evidence the
dependence of the potential on the spin state. Were parity preserved, this
would not be the result; however, by virtue of the explicit breaking of
parity, as induced by the Chern-Simons term, expressions (\ref{V1}), (\ref
{V2}), (\ref{V3}) differ from one another as it can be understood on the
basis of parity transformation arguments. Another signal of parity-breaking
is the linear dependence of $V$ on $l$: $l\rightarrow -l$\ is not a symmetry
of the potential.

Although the gauge invariance is broken by the appearance of a Proca mass
during the SSB, one expects that the interaction potential associated to the
system comes to preserve the characteristics of the original Lagrangian
(before the SSB). This fact leads us to study a way to assure the gauge
invariance of the effective interaction potential. Analysis of the Galilean
limit of the field theories in (1+2) dimensions, carried out by Hagen \cite
{Hagen}, have shown that the 2-body scattering problem, as mediated by a
gauge particle, must lead to an effective potential that preserves the
structure of a perfect square form $(l-\alpha ^{2})^{2}$, and can be
identified with the Aharonov-Bohm scattering potential. {\bf \ }The quartic
order term $\left( \alpha ^{4}\right) $ is related to the presence of
2-photon diagrams induced by the seagull vertex $\left( \varphi ^{\ast
}\varphi A_{\mu }A^{\mu }\right) $, and thus associated to the gauge
invariance of the resulting potential. In this way, the potential structure $%
(l-\alpha ^{2})^{2}$ must be also pursued in more complex electron-electron
scatterings panoramas, in order to ensure gauge invariance. Actually, this
is just the signal of a more general result. Electron-electron scatterings,
in general, no matter the complexity of the interactions, must exhibit the
combination $(l-\alpha ^{2})^{2}$ for the sake of gauge invariance of the
final result. This kind of procedure is found in Ref. \cite{Dobroliubov},
where a nonrelativistic interaction potential was derived in the context of
a MCS-QED$_{3}$\ (without scalar sector), in the perturbative regime, $%
1/k\ll 1,$ with $k$ being the statistic parameter \ (in our present case $%
k\equiv $ $4\pi \theta /e_{3}^{2})$. In this reference, in order to ensure
the gauge invariance, at the low-energy approximation, one takes into
account the two-photons diagrams, which amounts to adding up to the
tree-level potential the quartic order term $\left\{ \frac{e^{2}}{2\pi
\theta }[1-\theta rK_{1}(\theta r)]\right\} ^{2}$, turning out into the
following gauge-invariant effective potential form{\bf \ }\cite{Kogan},\cite
{Dobroliubov}{\bf :} 
\begin{equation}
V_{{\rm MCS}}(r)=\frac{e^{2}}{2\pi }\left[ 1-\frac{\theta }{m_{e}}\right]
K_{0}(\theta r)+\frac{1}{m_{e}r^{2}}\left\{ l-\frac{e^{2}}{2\pi \theta }%
[1-\theta rK_{1}(\theta r)]\right\} ^{2}~.  \label{Vmcs}
\end{equation}
In the expression above, the first term corresponds to the electromagnetic
potential, whereas the last one incorporates the centrifugal barrier $\left(
l/mr^{2}\right) ,$ the Aharonov-Bohm term and the 2-photon exchange term.
One observes that this procedure becomes necessary when the model is
analyzed or defined out of the pertubative limit. In Ref. \cite{Georgelin},
for instance, one accomplishes an evaluation of the scattering potential, in
the Born approximation, whose final result is not supplemented by the term $%
\left\{ \frac{e^{2}}{2\pi \theta }[1-\theta rK_{1}(\theta r)]\right\} ^{2}$,
under the justification that derivation has been done in the pertubative
regime $\left( 1/k\ll 1\right) .$ In such a regime, the 2-photon term
becomes negligible (for it is proportional to $1/k^{2})$ and shows itself
unable to jeopardize the gauge invariance of the model.

In a scenery where one searches for applications to Condensed Matter
Physics, one must require $\theta \ll m_{e}$, and the scattering potential
given by Eq.(\ref{Vmcs}) then comes out positive. This implication prevents
a possible application of this kind of model to superconductivity, where the
characteristic energies are of $meV$ order. Since the effective electron
mass ($m_{\text{{\tiny eff}}}=m_{e}+yv^{2})$ is $\sim 10^{5}eV,$ energy
scale much greater than that corresponding to the condensed matter
interactions $\left( meV\right) $, one must impose the following condition
on the physical excitations of the model: 
\begin{equation}
m_{\text{{\tiny eff}}}\gg \vartheta ,M_{A},M_{\pm }\text{ .}
\label{Condmat-lim}
\end{equation}
In the limit $M_{A}\rightarrow 0,$ one has: $M_{+}\sim \vartheta $; in this
situation, the dimensionless parameter ${e_{3}^{2}}/{M}_{+}\ $\ reduces to ${%
e_{3}^{2}}/{\vartheta ,}$ that now lies outside the pertubative regime,
since $\vartheta $ is now small $\left( \sim meV\right) $. Therefore, in
this energy scale, our results may not be restricted to the pertubative
limit; the consideration of the 2-photon term to Eqs.(\ref{V1}, \ref{V2}, 
\ref{V3}) becomes then relevant in order to assure the gauge invariance of
these potentials. As a final result, one rewrites the three expressions for
the effective-gauge-invariant scattering potentials: 
\begin{eqnarray}
V_{\text{eff}_{\uparrow \uparrow }}(r) &=&-{\frac{1}{2\pi }}%
2v^{2}y^{2}K_{0}(M_{H}r)+\frac{e_{3}^{2}}{2\pi }\biggl\{\left[ (C_{+}-\frac{C%
}{m_{\text{{\tiny eff}}}}M_{+}^{2}\right] K_{0}(M_{+}r)+\biggl[C_{-}+\frac{C%
}{m_{\text{{\tiny eff}}}}M_{-}^{2}\biggr]K_{0}(M_{-}r)\biggr\}  \nonumber \\
&&+\frac{1}{m_{\text{{\tiny eff}}}r^{2}}\left\{ l+\frac{e_{3}^{2}}{2\pi }%
Cr[M_{+}K_{1}(M_{+}r)-M_{-}K_{1}(M_{-}r)]\right\} ^{2}~,  \label{Veff1}
\end{eqnarray}
\begin{eqnarray}
V_{\text{eff}_{\uparrow \downarrow }}(r) &=&-{\frac{1}{2\pi }}%
2v^{2}y^{2}K_{0}(M_{H}r)+\frac{e_{3}^{2}}{2\pi }\text{ }\left[
C_{+}K_{0}(M_{+}r)+C_{-}K_{0}(M_{-}r)\right] +\frac{1}{m_{\text{{\tiny eff}}%
}r^{2}}\biggl \{l+\frac{e_{3}^{2}}{2\pi }Cr[M_{+}K_{1}(M_{+}r)+  \nonumber \\
&&-M_{-}K_{1}(M_{-}r)]\biggr\}^{2},  \label{Veff2}
\end{eqnarray}
\begin{align}
V_{\text{eff}_{\downarrow \downarrow }}(r)& =-{\frac{1}{2\pi }}%
2v^{2}y^{2}K_{0}(M_{H}r)+\frac{e_{3}^{2}}{2\pi }\biggl\{\left[ C_{+}+\frac{C%
}{m_{\text{{\tiny eff}}}}M_{+}^{2}\right] K_{0}(M_{+}r)+\biggl[C_{-}-\frac{C%
}{m_{\text{{\tiny eff}}}}M_{-}^{2}\biggr]K_{0}(M_{-}r)\biggr\}  \nonumber \\
& \text{ }+\frac{1}{m_{\text{{\tiny eff}}}r^{2}}\left\{ l+\frac{e_{3}^{2}}{%
2\pi }Cr[M_{+}K_{1}(M_{+}r)-M_{-}K_{1}(M_{-}r)]\right\} ^{2},  \label{Veff3}
\end{align}
where $\frac{l^{2}}{mr^{2}}$ represents the centrifugal barrier, and the
term proportional to $C^{2}$ comes from the 2-photon exchange.

In the energy scale given by condition (\ref{Condmat-lim}), the
proportionality coefficients of $V_{2}(r)$ become negligible: 
\begin{equation}
m_{\text{{\tiny eff}}}\gg \vartheta ,M_{A},M_{\pm }\text{ \ \ \ \ \ }%
\Longrightarrow \text{ \ \ }\frac{C}{m_{\text{{\tiny eff}}}}M_{+}^{2}\ll 1,%
\text{ }\frac{C}{m_{\text{{\tiny eff}}}}M_{-}^{2}\ll 1.
\label{approximation}
\end{equation}
As a consequence of these considerations, one can observe that only the
first term of the expressions \ (\ref{Veff1}, \ref{Veff2}, \ref{Veff3}) is
attractive, which corresponds to the Higgs interaction. At the same time,
the potential $V_{2}(r)$ reveals itself small before $V_{1}(r)$ and $%
V_{3}(r),$ leading to a simplification in the expressions $\left( \text{\ref
{Veff1}}\right) ,$ $\left( \text{\ref{Veff2}}\right) ,$ $\left( \text{\ref
{Veff3}}\right) $, that degenerate to a single form:

\begin{align}
V_{\text{eff}}(r)& =-{\frac{1}{2\pi }}2v^{2}y^{2}K_{0}(M_{H}r)+\frac{%
e_{3}^{2}}{2\pi }\text{ }\biggl[C_{+}K_{0}(M_{+}r)+C_{-}K_{0}(M_{-}r)\biggr]+%
\frac{1}{m_{\text{{\tiny eff}}}r^{2}}\biggl\{l+  \nonumber \\
+& \frac{e_{3}^{2}}{2\pi }Cr[M_{+}K_{1}(M_{+}r)-M_{-}K_{1}(M_{-}r)]\biggr\}%
^{2},  \label{Veffdegenerado}
\end{align}
The fact that $C_{\pm }>0,$ $\forall $ $\vartheta ,M_{A}$ makes the second
term (proportional to $e^{2}/2\pi )$ of the equation above to be positive,
revealing the repulsive nature of gauge sector. This trivial analysis shows
that the potentials $\left( \text{\ref{Veff1}}\right) ,$ $\left( \text{\ref
{Veff2}}\right) ,$ $\left( \text{\ref{Veff3}}\right) $ will be attractive
only when the contribution originated from the Yukawa interaction overcomes
the one coming from the gauge sector, which can be achieved by accomplishing
a suitable fitting on the model parameters. The fulfillment of this
condition can render the formation of $e^{-}e^{-}$ bound states feasible ,
once the above potentials are ``weak'' in the sense of Kato criterion,
analyzed by Chadan {\it et al.} \cite{Chadan} in the context of the
low-energy scattering theory in $(1+2)$ dimensions.

Finally, it is instructive to show how the gauge sectors of the potentials $%
\left( \text{\ref{Veff1}}\right) $, $\left( \text{\ref{Veff2}}\right) $,$%
\left( \text{\ref{Veff3}}\right) $ behave in the limit of a vanishing Proca
mass: $M_{A}\rightarrow 0$. In this case, the propagator of the gauge field
reduces to the Maxwell-Chern-Simons one, leading to the following limits:

\begin{equation}
M_{+}\longrightarrow\theta;M_{-}\longrightarrow0;C_{+}\longrightarrow
1;C_{-}\longrightarrow0;K_{1}(M_{-}r)\longrightarrow\frac{1}{M_{-}r}%
;C\longrightarrow\frac{1}{\theta};
\end{equation}

\begin{equation}
\lim_{M_{A}\longrightarrow 0}V_{_{\uparrow \uparrow }}(r)=\frac{e_{3}^{2}}{%
2\pi }(1-\frac{\theta }{m_{\text{{\tiny eff}}}})K_{0}(\theta r)+\frac{1}{m_{%
\text{{\tiny eff}}}r^{2}}\left[ l-\frac{e_{3}^{2}}{2\pi \theta }(1-\theta
rK_{1}(\theta r))\right] ^{2},  \label{limit1}
\end{equation}
\begin{equation}
\lim_{M_{A}\longrightarrow 0}V_{_{\uparrow \downarrow }}(r)=\frac{e_{3}^{2}}{%
2\pi }K_{0}(\theta r)+\frac{1}{m_{\text{{\tiny eff}}}r^{2}}\left[ l-\frac{%
e_{3}^{2}}{2\pi \theta }(1-\theta rK_{1}(\theta r))\right] ^{2},
\end{equation}
\begin{equation}
\lim_{M_{A}\longrightarrow 0}V_{_{\downarrow \downarrow }}(r)=\frac{e_{3}^{2}%
}{2\pi }(1+\frac{\theta }{m_{\text{{\tiny eff}}}})K_{0}(\theta r)+\frac{1}{%
m_{\text{{\tiny eff}}}r^{2}}\left[ l-\frac{e_{3}^{2}}{2\pi \theta }(1-\theta
rK_{1}(\theta r))\right] ^{2}.
\end{equation}
One remarks that Eq. (\ref{limit1}) encloses exactly the same result
achieved by Dobrolibov \cite{Dobroliubov} {\it et al.} and others \cite
{Kogan}, \cite{Groshev} for the scattering of two up-polarization electrons,
which enforces the generalization realized in this paper.

\section{Numerical Analysis}

The numerical procedure adopted here consists on the implementation of the
variational method for the Schr\"{o}dinger equation supplemented by the
interaction potential (\ref{Veffdegenerado}). In this sense, it is necessary
to expose some properties of the wavefunction representing the $e^{-}e^{-}$
and of the two-dimensional Schr\"{o}dinger equation.

\subsection{The composite wave-function and the Schr\"{o}dinger equation}

\label{sec3}The Pauli exclusion principle states the antisymmetric character
of the total two-electron wavefunction $(\Psi )$ with respect to an
electron-electron permutation: $\Psi ({\bf \rho }_{1},s_{1,}{\bf \rho }%
_{2},s_{2})=-\Psi ({\bf \rho }_{2},s_{2,}{\bf \rho }_{1},s_{1}).$ Assuming
that no significant spin-orbit interaction takes place, the function $\Psi $
can be split into three independent functions:\ $\Psi (\rho _{1},s_{1,}\rho
_{2},s_{2})=\psi ({\bf R})\varphi ({\bf r})\chi \left( s_{1},s_{2}\right) $,
which represent, respectively, the center-of-mass wave function, the
relative one, and the spin wave function (${\bf R}$ and $s$ being the
center-of-mass and spin coordinates respectively, while ${\bf r}$ is the
relative coordinate of the electrons). Taking into account the Pauli
principle, the total wavefunction $\Psi $ in the center-of-mass frame reads
as 
\begin{equation}
\Psi ^{S=1}=\varphi _{{\rm odd}}({\bf r})\chi _{{\rm even}%
}^{S=1}(s_{1},s_{2})~,\text{ \ \ }\Psi ^{S=0}=\varphi _{{\rm even}}({\bf r}%
)\chi _{{\rm odd}}^{S=0}(s_{1},s_{2})~,  \label{antisym2}
\end{equation}
where $\chi ^{S=0},$ $\chi ^{S=1}$, $\varphi _{{\rm even}}({\bf r}),$ $%
\varphi _{{\rm odd}}({\bf r})$ stand respectively for the (antisymmetric)
singlet spin-function, the (symmetric) spin triplet, the even space-function
($l=0$: $s$-wave, $l=2$: $d$-wave), and the odd space-function ($l=1$: $p$%
-wave: , $l=3$: $f$-wave).

Within the nonrelativistic approximation, the binding energy associated to
an $e^{-}e^{-}$ pair is given by planar Schr\"{o}dinger equation for the
relative space-function $\varphi (r),$ 
\begin{equation}
\frac{\partial ^{2}\varphi (r)}{\partial r^{2}}+\frac{1}{r}\frac{\partial
\varphi (r)}{\partial r}-\frac{l^{2}}{r^{2}}\varphi (r)+2\mu _{{\rm eff}%
}[E-V(r)]\varphi (r)=0~,  \label{diff1}
\end{equation}
where $V(r)$ represents the interaction potential given by Eq. (\ref
{Veffdegenerado}), and $\mu _{{\rm eff}}=\frac{1}{2}m_{\text{{\tiny eff}}},$
is the effective reduced mass of the system. By means of the following
transformation $\varphi (r)=\frac{1}{\sqrt{r}}~g(r)$, one has 
\begin{equation}
\frac{\partial ^{2}g(r)}{\partial r^{2}}-\frac{l^{2}-\frac{1}{4}}{r^{2}}%
g(r)+2\mu _{{\rm eff}}[E-V(r)]g(r)=0~.  \label{diff2}
\end{equation}

\subsection{The Variational Method and the Choice of the trial function}

To work out the variational method, one must take as starting point the
choice of the trial function that represents the generic features of the $%
e^{-}e^{-}$ pair. The definition of a trial function must observe some
conditions, such as the asymptotic behavior at infinity, the analysis of its
free version and its behavior at the origin. For a zero angular momentum ($%
l=0$) state, the Eq.(\ref{diff2}) becomes 
\begin{equation}
\biggl\{\frac{\partial ^{2}}{\partial r^{2}}+\frac{1}{4r^{2}}+2\mu _{{\rm eff%
}}[E+C_{s}K_{0}(M_{H}r)]\biggr\}g(r)=0,~  \label{diff3}
\end{equation}
whose free version ($V(r)=0)$, for $l=0$ state, $\left[ \frac{\partial ^{2}}{%
\partial r^{2}}+\frac{1}{4r^{2}}+k^{2}\right] u(r)=0~,$ has as solution $%
u(r)=B_{1}\sqrt{r}J_{0}(kr)+B_{2}\sqrt{r}Y_{0}(kr)$, with $B_{1}$ and $B_{2}$
being arbitrary constants and $k=\sqrt{2\mu _{{\rm eff}}E}$. In the limit $%
r\rightarrow 0$, this solution goes simply as $u(r)\longrightarrow \sqrt{r}%
+\lambda \sqrt{r}\ln (r).$ Since the second term in the last equation
behaves like an attractive potential, $-1/4r^{2}$, this implies the
possibility of obtaining a bound state ($E<0$) even for $V(r)=0$ \cite
{Chadan}. This is not physically acceptable, leading to a restriction on the
needed self-adjoint extension of the differential operator $%
-d^{2}/dr^{2}-1/4r^{2}$. Among the infinite number of self-adjoint
extensions of this operator, the only physical choice corresponds to the
Friedrichs extension ($B_{2}=0$), which behaves like $\sqrt{r}$ at the
origin, indicating this same behavior for $u(r)$. In this way the behavior
of the trial function at the origin is determined. The complete equation, $%
V(r)\neq 0$, will preserve the self-adjointness of free Hamiltonian, if the
potential is ``weak'' in the sense of the Kato condition: $\
\int_{0}^{\infty }r(1+|\ln (r)|)|V(r)|dr<\infty ~.$ Provided the interaction
potential, given by Eq. (\ref{Veffdegenerado}), satisfies the Kato
condition, the self-adjointness of the total Hamiltonian is assured and the
existence of bound states is allowed. On the other hand, at infinity, the
trial function must vanish asymptotically in order to fulfill square
integrability. Therefore, a good choice can then be given by $g(r)=f(r)\exp
(-\beta r),$ where $f(r)$ is a well-behaved function that satisfies the
limit condition: $\lim_{r\rightarrow 0}f(r)=\sqrt{r}$. By simplicity, the
trial function (for zero angular momentum) read as 
\begin{equation}
g(r)=\sqrt{r}\exp (-\beta r)~,  \label{funcaoteste1}
\end{equation}
where $\beta $ is a free parameter whose variation approximately determines
an energy minimum.

An analogous procedure can be undertaken to determine the behavior of the
trial function when the angular momentum is different from zero ($l\neq 0$).
In this case, and in the limit $r\rightarrow 0$, Eq.(\ref{diff2}) reduces to 
$\left[ \frac{\partial ^{2}}{\partial r^{2}}-\frac{l^{2}-\frac{1}{4}}{r^{2}}%
+k^{2}\right] u(r)=0~,$ whose general solution reads as $%
u(r)=B_{1}r^{(l+1/2)}+B_{2}r^{(-l+1/2)}$. For $l>0,$ the choice $r^{(l+1/2)}$
entails a trial function that is well-behaved at the origin. Since the
Schr\"{o}dinger equation depends only on $l^{2}$, any of the choices, $l>0$
or $l<0$, is enough to provide the energy values of the physical states and
one gets 
\begin{equation}
g(r)=r^{1/2+l}\exp (-\beta r)~,  \label{funcaoteste2}
\end{equation}
where $\beta $ is again a spanning free parameter to be numerically fixed in
order to maximize the binding energy. Though this last result is
mathematically correct, we should point out that the discussion regarding
non-zero angular momentum states here is merely for the sake of
completeness. The true wave-function in this case should include the angular
components which remain precluded in this approach.

\subsection{The Analysis of the Potential and the Numerical Data}

The numerical analysis of the potentials $V_{\text{eff}_{\uparrow \uparrow
}},V_{\text{eff}_{\uparrow \downarrow }},V_{\text{eff}_{\downarrow
\downarrow }}$ is \ totally dependent on the parameters of the
field-theoretical model. As a first step, it is convenient to realize an
analysis on the relevant parameters and thereafter to initiate a numerical
procedure. The central purpose of this section is to demonstrate that the
potentials obtained are attractive and lead to the formation of bound states 
$e^{-}e^{-}$, whose energy is situated into a range relevant to some
Condensed Matter systems, like the high-T$_{c}$ superconductors.

As well-known, to parallel-spin states (spin triplet) there must be a p-wave
(spin triplet and $l=1$) associated, whereas the antiparallel-spin states
(spin singlet) are linked to an s-wave (spin singlet and $l=0$). Here,
despite the parity-breakdown to be associated to the state $l=1,$ the s-wave
can also appear as solution, since the breakdown is not necessarily manifest
in all states. Given the degeneracy of the potentials $V_{\text{eff}%
_{\uparrow \uparrow }},V_{\text{eff}_{\uparrow \downarrow }},V_{\text{eff}%
_{\downarrow \downarrow }}$on the reduced potential (\ref{Veffdegenerado}),
the issue concerning the wavefunction symmetry looses some of its status:
both the s- and p-wave appear as solutions for the system. According to Eqs.
(\ref{funcaoteste1}), (\ref{funcaoteste2}), the implementation of the
variational method requires a trial-function with $r^{1/2}-$behaviour at the
origin in the case of an s-wave and a $r^{3/2}-$behaviour for a p-wave.

Before starting the numerical calculations, it is instructive to show the
relevant parameters:

\begin{align}
e_{3}^{2}& =\frac{e^{2}}{l_{\perp }}=\frac{1}{137,04}\frac{1973,26}{l_{\perp
}}=\frac{14,399}{l_{\perp }}, \\
\alpha & =\frac{\vartheta }{M_{A}}, \\
\zeta & <0,\text{ }\lambda \geq \frac{3}{4}\frac{|\zeta |}{\upsilon ^{2}},
\label{ineq} \\
\lambda & =\frac{3}{4}\frac{|\zeta |}{\nu ^{2}}\Longrightarrow M_{H}^{2}=\nu
^{2}|\zeta |,  \label{Mhiggs4} \\
\lambda & =\frac{|\zeta |}{\nu ^{2}}\Longrightarrow M_{H}^{2}=2\nu
^{2}|\zeta |.  \label{Mhiggs5}
\end{align}

Specifically, in $D=1+2$, the electromagnetic coupling constant squared, $%
e_{3}^{2}$, has dimension of mass, rather than the dimensionless character
of the usual four-dimensional QED$_{4}$ coupling constant. This fact might
be understood as a memory of the third dimension that appears (into the
coupling constant) when one tries to work with a theory intrinsically
defined in three space-time dimensions. This dimensional peculiarity could
be better implemented through the definition of a new coupling constant in
three space-time dimensions \cite{Kogan},\cite{Randjbar}: $e\rightarrow
e_{3}=e/\sqrt{l_{\perp }}$, where $l_{\perp }$ represents a length
orthogonal to the planar dimension. The smaller is $l_{\perp }$, the smaller
is the remnant of the frozen dimension, the larger is the planar character
of the model and the coupling constant $e_{3}$, what reveals its effective
nature. In this sense, it is instructive to notice that the effective value
of $e_{3}^{2}$\ is larger than $e^{2}=1/137$\ whenever $l_{\perp }$\ $%
<1973.26$\ \AA , since 1 (\AA )$^{-1}=1973.26$\ $eV$. This particularity
broadens the repulsive interaction for small $l_{\perp }$ and requires an
even stronger Higgs contribution to account for a total attractive
interaction. Finally, this parameter must be evaluated inside a range
appropriated to not jeopardize the planar nature of the system, so that one
requires that:\ $2<l_{\perp }<15$\AA . The parameter $\alpha $ is defined as
the ratio between the Proca mass and the Chern-Simons mass, while $\zeta
,\lambda $ are parameters of $V-$potential and are important to assure a
stable vacuum, condition given by Eq. (\ref{ineq}). The imposition of some
relations between $\zeta ,\lambda ,\nu ^{2}$, like Eqs.(\ref{Mhiggs4}) e (%
\ref{Mhiggs5}), imply a kind of expression for the Higgs mass that exhibit
dependence only on $\nu ^{2}$ and $|\zeta |$. This set of conditions impose
a lower bound for the Higgs mass: $M_{H\min }^{2}=3|\zeta |/4\lambda $.

Besides the factors above, the entire determination of the potential (\ref
{Veffdegenerado}) also depends on $v^{2},$ the vacuum expectation value
(v.e.v.), and on $y$, the parameter that measures the coupling between the
fermions and the Higgs scalar. Being a free parameter, $v^{2}$ indicates the
energy scale of the spontaneous breakdown of the $U(1)-$local symmetry,
usually determined by some experimental data associated to the phenomenology
of the model under investigation, as it occurs in the electroweak
Weinberg-Salam model, for example. On the other hand, the $y-$parameter
measures the coupling between the fermions and the Higgs scalar, working in
fact as an effective constant that embodies contributions\ of all possible
mechanisms of the electronic interaction via Higgs-type (scalar)
excitations, as the spinless bosonic interaction mechanisms: phonons,
plasmons, and other collective excitations. This theoretical similarity
suggests an identification of the field theory parameter with an effective
electron-scalar coupling (instead of an electron-phonon one): $y\rightarrow
\lambda _{{\rm es}}$.

The numerical analysis is developed by means of the implementation of the
variational method on the Schr\"{o}dinger equation, supplemented by the
degenerated potential. The procedure is initiated by the use of the an
s-wave trial function: $g\left( r\right) =r^{1/2}e^{-\beta r}$, given by Eq. 
$\left( \text{\ref{funcaoteste1}}\right) $. Tables \ref{table1} and \ref
{table2} exhibit the values of the $e^{-}e^{-}$ bound state and the average
length of the $e^{-}e^{-}$ state ($\xi _{ab})$ for $V_{\text{eff}},$ in
accordance with the input parameters ($\nu ^{2},Z,\alpha ,y,\zeta )$, for $%
l=0.$\ The degenerated potential obviously assures the following equality: 
{\small E}$_{ee\uparrow \uparrow }=${\small E}$_{ee\downarrow \downarrow }=$%
{\small E}$_{ee\uparrow \downarrow },$ $\xi _{ab\uparrow \uparrow }=\xi
_{ab\downarrow \downarrow }=\xi _{ab\uparrow \downarrow }.$ Table \ref
{table3} contains numerical data generated by the variational method, for $%
l=1,$ starting from the following trial function: $\varphi \left( r\right)
=r^{3/2}e^{-\beta r},$ given by Eq. (\ref{funcaoteste2}).

\qquad

\begin{table}[h]
\caption{Input parameters: $\protect\nu ^{2},l_{\perp },\protect\alpha ,%
\protect\zeta ,M_{H}^{2}=\protect\nu ^{2}|\protect\zeta |$ and $l=0$; output
numerical data: $E_{e^{-}e^{-}}\ $and $\protect\xi _{ab}$. Trial Function: $%
\protect\varphi \left( r\right) =r^{1/2}e^{-\protect\beta r}$\ \ \ \ \ \ \ \
\ \ \ \ \ \ \ \ \ \ \ \ \ \ \ \ \ \ \ \ \ \ \ \ \ \ \ \ \ \ }
\label{table1}
\ 
\begin{tabular}{|l|l|l|l|l|l|l|l|l|}
\hline
$v^{2}$(meV) & $l_{\perp }$(\AA ) & $y$ & \ \ $\alpha $ & $\zeta $ (eV) & $%
M_{H}=\sqrt{\nu ^{2}|\zeta |}$ & $\beta $ & $E_{e^{-}e^{-}}$(meV)$\ $ & $\xi
_{ab}$(\AA ) \\ \hline
&  &  &  &  &  &  &  &  \\ \hline
$47.0$ & $10.0$ & $4.0$ & $1.0$ & $4.0$ & $433.0$ & $51.1$ & $-59.2$ & $19.3$
\\ \hline
$47.0$ & $10.0$ & $4.0$ & $0.5$ & $4.0$ & $433.0$ & $51.8$ & $-23.7$ & $19.0$
\\ \hline
$48.0$ & $10.0$ & $4.0$ & $0.5$ & $4.0$ & $438.0$ & $29.8$ & $-50.2$ & $16.6$
\\ \hline
$48.0$ & $10.0$ & $3.9$ & $1.0$ & $4.0$ & $438.0$ & $29.8$ & $-24.8$ & $33.1$
\\ \hline
$60.0$ & $8.0$ & $4.0$ & $1.0$ & $8.0$ & $693.0$ & $71.1$ & $-33.3$ & $13.9$
\\ \hline
$60.0$ & $8.0$ & $4.0$ & $0.5$ & $6.0$ & $600.0$ & $69.2$ & $-32.8$ & $14.3$
\\ \hline
$60.0$ & $8.0$ & $3.9$ & $1.0$ & $5.0$ & $548.0$ & $27.1$ & $-30.4$ & $36.4$
\\ \hline
$70.0$ & $7.0$ & $4.0$ & $0.4$ & $7.0$ & $700.0$ & $89.2$ & $-62.7$ & $11.6$
\\ \hline
$70.0$ & $7.0$ & $4.0$ & $0.6$ & $8.0$ & $748.0$ & $87.5$ & $-54.0$ & $11.3$
\\ \hline
$70.0$ & $7.0$ & $3.9$ & $1.0$ & $7.0$ & $700.0$ & $51.2$ & $-32.3$ & $19.3$
\\ \hline
$70.0$ & $7.0$ & $3.9$ & $0.5$ & $5.0$ & $590.0$ & $50.8$ & $-38.5$ & $19.4$
\\ \hline
\end{tabular}
\end{table}
\begin{table}[h]
\caption{ Input parameters: $\protect\nu ^{2},l_{\perp },\protect\alpha ,%
\protect\zeta ,M_{H}^{2}=\protect\nu ^{2}|\protect\zeta |$ and $l=0$; output
numerical data: $E_{e^{-}e^{-}}\ $and $\protect\xi _{ab}$. Trial Function: $%
\protect\varphi \left( r\right) =r^{1/2}e^{-\protect\beta r}$\ \ \ \ \ \ \ \
\ \ \ \ \ \ \ \ \ \ \ \ \ \ \ \ \ \ \ \ \ \ \ \ \ \ \ \ \ \ }
\label{table2}
\begin{tabular}{|l|l|l|l|l|l|l|l|l|}
\hline
$v^{2}$ (meV) & $l_{\perp }$(\AA ) & $y$ & \ \ $\alpha $ & $\zeta $ (eV) & $%
M_{H}=\sqrt{2\nu ^{2}|\zeta |}$ & $\beta $ & $E_{e^{-}e^{-}}$(meV)$\ $ & $%
\xi _{ab}$(\AA ) \\ \hline
&  &  &  &  &  &  &  &  \\ \hline
$40.0$ & $12.0$ & $4.0$ & $1.0$ & $2.0$ & $400.0$ & $56.1$ & $-54.1$ & $17.6$
\\ \hline
$40.0$ & $12.0$ & $4.0$ & $0.5$ & $2.0$ & $400.0$ & $59.2$ & $-24.5$ & $16.7$
\\ \hline
$40.0$ & $12.0$ & $4.0$ & $0.3$ & $2.0$ & $400.0$ & $58.1$ & $-17.2$ & $17.0$
\\ \hline
$40.0$ & $12.0$ & $4.0$ & $1.0$ & $2.5$ & $447.2$ & $57.9$ & $-31.4$ & $17.0$
\\ \hline
$50.0$ & $10.0$ & $4.0$ & $1.5$ & $6.3$ & $793.7$ & $79.1$ & $-41.1$ & $12.5$
\\ \hline
$50.0$ & $10$ & $4.0$ & $1.5$ & $5.3$ & $728.0$ & $79.1$ & $-63.1$ & $12.5$
\\ \hline
$60.0$ & $8.0$ & $4.0$ & $0.5$ & $3.0$ & $600.0$ & $69.2$ & $-32.8$ & $14.3$
\\ \hline
$60.0$ & $8.0$ & $3.9$ & $0.1$ & $2.0$ & $489.9$ & $51.2$ & $-38.6$ & $19.3$
\\ \hline
$60.0$ & $8.0$ & $3.9$ & $1.0$ & $2.0$ & $489.9$ & $27.2$ & $-62.8$ & $36.3$
\\ \hline
$80.0$ & $6.0$ & $4.0$ & $0.5$ & $4.0$ & $800.0$ & $79.1$ & $-40.2$ & $12.5$
\\ \hline
$80.0$ & $6.0$ & $4.0$ & $0.1$ & $3.0$ & $692.8$ & $78.1$ & $-76.7$ & $12.6$
\\ \hline
$80.0$ & $6.0$ & $3.9$ & $0.5$ & $2.5$ & $632.5$ & $27.1$ & $-36.0$ & $36.4$
\\ \hline
$80.0$ & $6.0$ & $3.9$ & $0.6$ & $2.5$ & $632.5$ & $29.8$ & $-45.7$ & $33.1$
\\ \hline
\end{tabular}
%
%
%
%
%
%
%
%
%
%
%
%
%
%
%
%
%
%
%
%
%
%
%
%
%
%
%
%
%
%
%
%
%
%
%
%
%
%
%
%
%
%
%
%
%
%
%
%
%
%
%
%
\end{table}

\qquad \qquad 
\begin{table}[h]
\caption{Input parameters: $\protect\nu ^{2},l_{\perp },\protect\alpha ,%
\protect\zeta ,M_{H}^{2}=2\protect\nu ^{2}|\protect\zeta |$ and $l=1$;
output data: $E_{e^{-}e^{-}}\ $and $\protect\xi _{ab}$. \ Trial function: $%
\protect\varphi \left( r\right) =r^{3/2}e^{-\protect\beta r}$}
\label{table3}
\begin{tabular}{|l|l|l|l|l|l|l|l|l|}
\hline
$v^{2}$(meV) & $l_{\perp }$(\AA ) & $y$ & \ $\alpha $ & $\zeta $ (eV)\  & $%
M_{H}=\sqrt{2\nu ^{2}|\zeta |}$ & $\beta $ & $E_{e^{-}e^{-}}$ (meV) & $\xi
_{ab}$(\AA )$\ $ \\ \hline
&  &  &  &  &  &  &  &  \\ \hline
$30.0$ & $16.0$ & $4.0$ & $2.0$ & $-2.0$ & $489.9$ & $55.1$ & $-71.5$ & $%
53.7 $ \\ \hline
$30.0$ & $15.5$ & $4.0$ & $2.0$ & $-3.0$ & $489.9$ & $40.7$ & $-23.2$ & $%
72.7 $ \\ \hline
$30.0$ & $15.5$ & $4.0$ & $3.0$ & $-4.0$ & $489.9$ & $42.2$ & $-56.2$ & $%
70.1 $ \\ \hline
$32.0$ & $15.0$ & $4.0$ & $2.0$ & $-3.0$ & $438.2$ & $70.7$ & $-49.5$ & $%
41.9 $ \\ \hline
$32.0$ & $15.0$ & $4.0$ & $1.0$ & $-2.0$ & $357.8$ & $51.1$ & $-18.0$ & $%
58.9 $ \\ \hline
$50.0$ & $10.0$ & $4.0$ & $1.5$ & $-5.3$ & $728.0$ & $80.9$ & $-43.9$ & $%
36.6 $ \\ \hline
$50.0$ & $10.0$ & $4.0$ & $1.5$ & $-4.0$ & $632.4$ & $79.1$ & $-77.3$ & $%
37.4 $ \\ \hline
$50.0$ & $10.0$ & $4.0$ & $0.8$ & $-3.0$ & $547.7$ & $72.4$ & $-49.5$ & $%
40.9 $ \\ \hline
$50.0$ & $10.0$ & $4.0$ & $0.5$ & $-3.0$ & $547.7$ & $42.9$ & $-25.0$ & $%
45.0 $ \\ \hline
$80.0$ & $6.5$ & $3.8$ & $1.0$ & $-4.0$ & $800.0$ & $61.3$ & $-21.6$ & $48.3$
\\ \hline
$80.0$ & $6.5$ & $3.8$ & $0.5$ & $-3.0$ & $692.8$ & $50.7$ & $-18.8$ & $58.4$
\\ \hline
$80.0$ & $6.5$ & $3.8$ & $0.5$ & $-2.5$ & $632.5$ & $51.8$ & $-52.3$ & $57.1$
\\ \hline
\end{tabular}
%
%
%
%
%
%
%
%
%
%
%
%
%
%
%
%
%
%
%
%
%
%
%
%
%
%
%
%
%
%
%
%
%
%
%
%
%
%
%
%
%
%
%
%
%
%
%
%
%
%
%
%
\end{table}

\ 

From the data of the Tables \ref{table1}, \ref{table2}, \ref{table3}, it is
possible to get an understanding of the influence of the parameters on the
values of the $e^{-}e^{-}$ energy and $\xi _{ab}$. When $|\zeta |$ and $\nu
^{2}$ increase, the Higgs mass grows up, reducing the range of the
attractive interaction, which is noticed through reduction of the bound
state energy. In the same way, the rising of the $\alpha -$parameter implies
a larger Chern-Simons mass and a reduction of the repulsive interaction
range, determining an increment of the bound state energy. The parameter $%
l_{\perp }$ acts directly in the coupling constant $e_{3}$: the bigger is $%
l_{\perp }$, the smaller is gauge coupling, and the smaller the repulsive
interaction, favoring again the increase of bound state energy. The
parameters $\nu ^{2}$ and $y$ act on the Higgs interaction coupling, in such
a way to promote a sensitive raising of the biding energy. In the particular
case of \ Table \ref{table3}, it is evident a sensitive enhancement in the
value of $\xi _{ab}$, a consequence of the isotropic trial function that
behaves as $r^{3/2}$ at the origin . This isotropic character results in a
non-realistic approximation, since the angular momentum state $l=1$\ must
exhibit some anisotropy. This observation attributes to the data of Table 
\ref{table3} a more qualitative aspect without invalidating the fundamental
result of this section: by means of a suitable fitting of the parameters, it
is possible to obtain values of the energy{\bf \ }and the correlation length
for the pairs $e^{-}e^{-}$\ \ inside a scale usual for some solid state
systems.

\section{General Conclusions}

The electron-electron interaction potentials, derived from a MCS
Electrodynamics with spontaneous symmetry breaking, puts in evidence the
physical possibility of electronic pairing and the formation of bound
states. This theoretical prediction occurs when the parameters of the model
are so chosen that the contribution stemming from the scalar (Higgs) sector
overcomes the contribution induced by the gauge boson exchange (always
repulsive in the energy scale relevant for the solid state excitations, $%
\theta \ll m_{e}$). The numerical results, displayed in Tables \ref{table1}, 
\ref{table2} and \ref{table3}, reveal the achievement of binding energies in
the $meV-$scale, and correlation lengths in the scale $10-30$\AA , which \
is a possible argument in favour of the MCS QED$_{3}$ adopted here to
address the electronic pairing process in the realm of some Condensed Matter
planar systems, with manifestation of parity-breaking, such as the Hall
systems (there are also some references that discuss the nonconservation of
parity symmetry in the context of the high-T$_{c}$ superconductors \cite
{Parity-breaking}).

Finally, we must observe that the present MCS model bypasses the
difficulties found by several other models \cite{Kogan}, \cite{Girotti}, 
\cite{Dobroliubov}, \cite{Groshev} that attempted to obtain $e^{-}e^{-}$
bound states considering only the exchange of vector bosons. The $v^{2}-$%
values disposed in Tables \ref{table1}, \ref{table2}, \ref{table3} reconfirm
the energy scale $\left( 10-100meV\right) $ for the breaking of U(1)-local
symmetry obtained in the framework of planar superconductors \cite{Bound}, 
\cite{Tese} and in the case of a parity-preserving electronic pairing \cite
{Tese}, \cite{Bound2}.

\section{Appendix}

In this Appendix one presents the spinor algebra $\ so$(1,2) that generates
the Dirac spinors, solutions of the Dirac equation in $D=1+2$ $\ $%
dimensions. The adopted metric is $\eta ^{\mu \nu }=(+,-,-),$ and the Dirac
equation is written as:\ 

\ 
\begin{align}
\left( \rlap{\hbox{$\mskip1 mu /$}}p-m\right) u_{+}(p)& =0,
\label{solution1} \\
\left( \rlap{\hbox{$\mskip1 mu /$}}p+m\right) u_{-}(p)& =0,
\label{solution2}
\end{align}
where $u_{+}(p),$ $u_{-}(p)$ stands for the positive energy spinors with
polarization ``up'' and ``down'' respectively. The solution of the equations
(\ref{solution1},\ref{solution2}) are given by:\ 
\begin{align}
u_{+}(p)& =\frac{\rlap{\hbox{$\mskip1 mu /$}}p+m}{\sqrt{2m(E+m)}}u_{+}(m,%
\overrightarrow{0}), \\
u_{-}(p)& =\frac{\rlap{\hbox{$\mskip1 mu /$}}p-m}{\sqrt{2m(E+m)}}u_{+}(m,%
\overrightarrow{0}),
\end{align}
where $u_{+}(m,\overrightarrow{0})$ and $u_{-}(m,\overrightarrow{0})$
represent an up-electron and down-electron (respectively) in the rest frame: 
\begin{equation}
u_{+}(m,\overrightarrow{0})=\left[ 
\begin{array}{c}
1 \\ 
0
\end{array}
\right] ;\text{ \ \ \ \ \ \ }u_{-}(m,\overrightarrow{0})=\left[ 
\begin{array}{c}
0 \\ 
1
\end{array}
\right]
\end{equation}

In $D=1+2,$ the generators of the group SO(1,2) are given by:

\begin{equation}
\Sigma ^{jl}=\frac{1}{4}[\gamma ^{j,}\gamma ^{l}],
\end{equation}
where the $\gamma $ matrices must satisfy the $so(1,2)$ algebra 
\begin{equation}
\lbrack \gamma _{\mu },\gamma _{\nu }]=2i\epsilon _{\mu \nu \alpha }\gamma
^{\alpha },
\end{equation}
and are taken by: $\gamma ^{\mu }=(\sigma _{z},-i\sigma _{x},i\sigma _{y}).$

Using this convention, the spinors $u_{+}(p),$ $u_{-}(p)$ are written at the
form:\ 

\begin{align}
u_{+}(p)& =\frac{1}{\sqrt{2m(E+m)}}\left[ 
\begin{array}{c}
E+m \\ 
-ip_{x}-p_{y}
\end{array}
\right] ;\overline{u}_{+}(p)=\frac{1}{\sqrt{2m(E+m)}}\left[ 
\begin{array}{cc}
E+m & -ip_{x}+p_{y}
\end{array}
\right] , \\
u_{-}(p)& =\frac{1}{\sqrt{2m(E+m)}}\left[ 
\begin{array}{c}
ip_{x}-p_{y} \\ 
E+m
\end{array}
\right] ;\overline{u}_{-}(p)=\frac{1}{\sqrt{2m(E+m)}}\left[ 
\begin{array}{cc}
-ip_{x}-p_{y} & E+m
\end{array}
\right] ,
\end{align}
They obviously satisfy the normalization condition: : $\overline{u}%
_{+}(p)u_{+}(p)=1$ and $\overline{u}_{-}(p)u_{-}(p)=-1.$

In the center of mass frame, the 3-momenta of the scattered electrons
(elastic scattering hypothesis) can be written as:

\begin{align*}
p_{1}& =(E,p,0),\text{ \ \ }p_{1}^{^{\prime }}=(E,p\cos \phi ,p\sin \phi ),
\\
p_{2}& =(E,-p,0),\text{ \ \ }p_{2}^{^{\prime }}=(E,-p\cos \phi ,-p\sin \phi
), \\
k& =\text{\ }p_{1}^{^{\prime }}-p_{1}=(0,p(\cos \phi -1),p\sin \phi ),
\end{align*}
where $\phi $ is the angle defined (in relation to the initial direction) by
the particles after the scattering.

Adopting this convention, the current terms are evaluated: \ 
\begin{align}
\left[ \overline{u}_{+}(p_{1}^{^{\prime }})\gamma _{_{0}}u_{+}(p_{1})\right]
& =\frac{(E+m)^{2}+p^{2}e^{i\theta }}{2m(E+m)}=\left[ \overline{u}%
_{+}(p_{2}^{^{\prime }})\gamma _{_{0}}u_{+}(p_{2})\right] ;\text{ \ \ \ } \\
\left[ \overline{u}_{+}(p_{1}^{^{\prime }})\gamma _{_{1}}u_{+}(p_{1})\right]
& =-\frac{p}{2m}(1+e^{i\theta })=-\left[ \overline{u}_{+}(p_{2}^{^{\prime
}})\gamma _{_{1}}u_{+}(p_{2})\right] ; \\
\left[ \overline{u}_{+}(p_{1}^{^{\prime }})\gamma _{_{2}}u_{+}(p_{1})\right]
& =\frac{-ip}{2m}(1-e^{i\theta })=-\left[ \overline{u}_{+}(p_{2}^{^{\prime
}})\gamma _{_{2}}u_{+}(p_{2})\right] ;
\end{align}

\begin{align}
\left[ \text{\ }\overline{u}_{-}(p_{1}^{^{\prime }})\gamma
_{_{0}}u_{-}(p_{1})\right] & =\frac{(E+m)^{2}+p^{2}e^{-i\theta }}{2m(E+m)}=%
\left[ \overline{u}_{-}(p_{2}^{^{\prime }})\gamma _{_{0}}u_{-}(p_{2})\right]
; \\
\left[ \overline{u}_{-}(p_{1}^{^{\prime }})\gamma _{_{1}}u_{-}(p_{1})\right]
& =-\frac{p}{2m}(1+e^{-i\theta })=-\left[ \overline{u}_{-}(p_{2}^{^{\prime
}})\gamma _{_{1}}u_{-}(p_{2})\right] ; \\
\left[ \overline{u}_{-}(p_{1}^{^{\prime }})\gamma _{_{2}}u_{-}(p_{1})\right]
& =\frac{ip}{2m}(1-e^{-i\theta })=-\left[ \overline{u}_{-}(p_{2}^{^{\prime
}})\gamma _{_{2}}u_{-}(p_{2})\right]
\end{align}
These current terms were used in the evaluation of the scattering amplitudes
in the nonrelativistic approximation: $p^{2}\ll m^{2}.$ Finally, given the
correlation between mass and spin \cite{Binegar}, valid in QED$_{3}$, it is
reasonable to inquire if the spinor $u_{-}(p)$ does not represent an
antiparticle rather than the spin-down particle. This issue is solved in the
Appendix of Ref. \cite{N.Cimento}, where one shows that the charge of the
spinor $u_{-}(p)\ $is equal to one of the spinor $u_{+}(p).$

\bigskip

{\bf Acknowledgments: }

M.M.F. Jr. is grateful to CCP-CBPF for the kind hospitality. J. A.
Helay\"{e}l-Neto expresses his gratitude to CNPq for the invaluable
financial help.

\end{document}